\documentclass[manuscript]{emulateapj}
\usepackage{times}

\slugcomment{}

\shorttitle{Mapping the Galactic Halo}
\shortauthors{De Propris et al. }

\begin{document}


\title{Mapping the Galactic halo with Blue Horizontal Branch stars
from the 2dF Quasar Redshift Survey}


\author{Roberto De Propris\altaffilmark{1}, Craig D. Harrison\altaffilmark{1}
\and Peter J. Mares\altaffilmark{2}}

\altaffiltext{1}{Cerro Tololo Inter-American Observatory, La Serena, Chile}
\altaffiltext{2}{Department of Astronomy, Cornell University, Ithaca, NY, USA}



\begin{abstract}

We use 666 blue horizontal branch (BHB) stars from the 2Qz redshift survey
to map the Galactic halo in four dimensions (position, distance and velocity).
We find that the halo extends to at least 100 kpc in Galactocentric distance,
and obeys a single power-law density profile of index $\sim -2.5$ in two
different directions separated by about 150$^{\circ}$ on the sky. This suggests
that the halo is spherical. Our map shows no large kinematically coherent
structures (streams, clouds or plumes) and appears homogeneous. However,
we find that at least 20\% of the stars in the halo reside in substructures
and that these substructures are dynamically young. The velocity dispersion
profile of the halo appears to increase towards large radii while the
stellar velocity distribution is non Gaussian beyond 60 kpc. We argue
that the outer halo consists of a multitude of low luminosity overlapping
tidal streams from recently accreted objects.

\end{abstract}


\keywords{Galaxy: formation --- Galaxy: halo --- stars: horizontal-branch}



\section{Introduction}

The motion of old stars preserves the fossil record of the earliest phases of
formation in the Milky Way, as was realized in the two seminal papers by Eggen,
Lynden-Bell \& Sandage (1962) and \cite{searle78}. The picture that has since 
emerged is one where the Galaxy has been built via a series of accretions and 
mergers (e.g., \citealt{freeman02}), in agreement with hierarchical structure 
formation scenarios \citep{johnston08,cooper09}. Consistent with these theoretical 
expectations, the halo is thickly populated by debris from past and on-going 
accretion events (e.g., Ibata, Gilmore \& Irwin 1994; Belokurov et al. 2006). 
Prominent debris features have also been observed in the halos of M31 \citep{ibata01,
ibata07,mcconnachie09}, M33 \citep{ibata07,mcconnachie09} and in other 
nearby spirals \citep{martinez10}.

At the same time, the inner halo (at galactocentric distance $R_{GC} < 20$ kpc)
of the Milky Way contains an old, dynamically smooth and metal-poor component, 
which was probably formed by rapid early merging or violent relaxation \citep{carollo07,
carollo09,bell08}. While the buildup of most of the inner halo is
believed to take place rapidly, the outer regions of the Milky Way should grow
more slowly via the disruption of dwarf galaxies: the outer halo is therefore
expected to be quite inhomogeneous and dominated by large streams, clouds and
plumes from infalling objects \citep{johnston08,cooper09}. The main question we
need to address is the relative role of mergers and accretion vs. {\it in situ}
formation, i.e., whether the observed structures represent the main mechanism
by which the Galaxy was formed or are only a comparatively minor contribution
over an ancient stellar component.

In order to clarify these issues we explore the space distribution and kinematics
of the outer halo using Blue Horizontal Branch (BHB) stars. These objects have
often been used as tracers of the Milky Way halo (e.g., \citealt{yanny00,xue08,
brown10}). BHB stars are comparatively bright and reliable standard candles,
are more common than other commonly used probes (such as Carbon stars or RR
Lyrae) and are representative of the {\it old} and {\it metal poor} stellar
population that constitutes the Galactic halo. These stars can be easily
identified spectroscopically, even at comparatively moderate resolution, but 
are sufficiently rare that extensive radial velocity surveys are needed to obtain 
a significant sample \citep{yanny00,xue08,brown10}.

Here we exploit a large suite of archival spectroscopy from the 2dF Quasar Redshift
Survey (2Qz -- \citealt{croom04}) to identify a sample of 666 BHB stars in the halo
out to 100 kpc and explore its structure and kinematics. We use this dataset to set 
limits to the size of the Milky Way halo, its shape, presence of streams, degree of 
substructure and the velocity dispersion profile out to large radii. The next section
describes the 2Qz survey and how we identified BHB stars. Section 3 discusses 
the 4D map of the Galactic halo we produce from these data. Section 4 examines
the question of substructure in the halo and the final section analyzes the halo
kinematics and discusses the results in the context of galaxy formation models,
especially as apply to the Milky Way.

\section{The 2Qz data: identification of BHB stars}

The data for this project come from archival observations of the 2Qz survey. The
2Qz obtained spectra for $\sim 50,000$ A-colored point sources with $15 < b_J <
20.9$ selected from Supercosmos survey photometry \citep{hambly01}. The targets lie 
in two $75^{\circ} \times 5^{\circ}$ strips on the sky, one on the celestial equator 
between 09h 30m $< RA <$ 14h 30m and $-2.5^{\circ} < \delta < +2.5^{\circ}$ and the 
other centred in  the Southern Galactic Cap between 22h 00m $< RA <$ 03h 00m and 
$-32.5^{\circ} <  \delta < -27.5^{\circ}$. We refer to these as the Northern and Southern 
samples, respectively. 

The targets were selected on the basis of their $u$, $b_J$ and $r_F$ colors
(photographic photometry from the SuperCosmos survey -- \citealt{hambly01})
to lie in the parameter space covered by known quasars (see \citealt{croom04}
for further details): targets were selected to have $b_J < 20.9$ and $-2.5 \leq
u-b_J \leq 1.5$ and $-1.5 \leq b_J-r_F \leq 2.5$, excluding a box with colors
$0.4 \leq u-b_J \leq 1.5$ and $0 \leq b_J-r_F \leq 2.5$ which contains main 
sequence stars. Spectroscopy for these objects was obtained during 1998-2003 with 
the Anglo-Australian 3.9m Telescope in Coonabarabran, NSW, Australia, using the 2 
degree field (2dF) multi-object spectrograph \citep{lewis02}. The spectra cover 
the entire optical range ($3500 < \lambda < 7500$ \AA) at a resolution 
of about 2000, with exposure times of at least one hour per target, although a 
fraction of the objects were exposed for considerably longer, if they happened to lie 
in a region where the 2dF tiles used by the survey overlapped (to insure greater
completeness) or the field was reobserved (because of lower than expected
signal).

As with all color-selected surveys, 2Qz suffers from significant contamination
from QSO-colored stellar objects, such as white dwarfs, disk A stars,
blue stragglers and BHB stars, although this is minimized by observing at
high galactic latitude. The data were reduced and redshifted via a semi-automated
technique and whenever a stellar redshift ($z=0$) was obtained, the object was 
discarded from further analysis but placed in the database. We proceeded to retrieve 
all stellar spectra from the database and classify them on the basis of the
equivalent widths of the H$_{\gamma}$ and H$_{\delta}$ Balmer lines, to
identify a sample of 666 {\it bona fide} BHB stars. Following \cite{yanny00,
xue08,brown10} we first measured radial velocities for all star by cross
correlating with synthetic templates from the library of \cite{munari05}: the
templates had temperatures and gravities typical of A-stars and field horizontal
branch stars \citep{beers01}. We then fit Gaussian curves to the H$\gamma$ and
H$\delta$ lines and measured the width of the Gaussian fit at 20\% of the normalized
continuum level: this indicator $D_{0.2}$ has been shown to be a good discriminator
between BHB stars, blue stragglers and other contaminants \citep{pier83,yanny00}. We 
only used spectra that we deemed to be of sufficient quality to allow a secure 
classification. In order to be included in our sample radial velocities had to
be determined to within 50 km s$^{-1}$ and the H$\gamma$ and H$\delta$ widths
had to have errors of less than 20\%. In order to classify stars as {\it bona
fide} BHB stars we require that the mean $D_{0.2}$, from both lines, lie between
17 and 31 \AA\ (as in \citealt{pier83,yanny00}, leaving a total of 666 BHB stars
in our sample. 

Figure 1 shows the distribution of the stars in the $u-g$ vs. $g-z$ plane, where
we transformed our Supercosmos $u-b_J$ and $b_J-r_F$ colors to the Sloan system 
by using stars in common in the equatorial region shared by 2Qz and the SDSS. 
This is at least qualitatively similar to the color distribution of stars in previous
work (e.g., compare Fig.~1 in \citealt{brown10}) and suggests that our sample
is comparable to those used in previous studies, in terms of selection criteria
and degree of contamination from blue stragglers and other A-colored objects.

\begin{figure}
\plotone{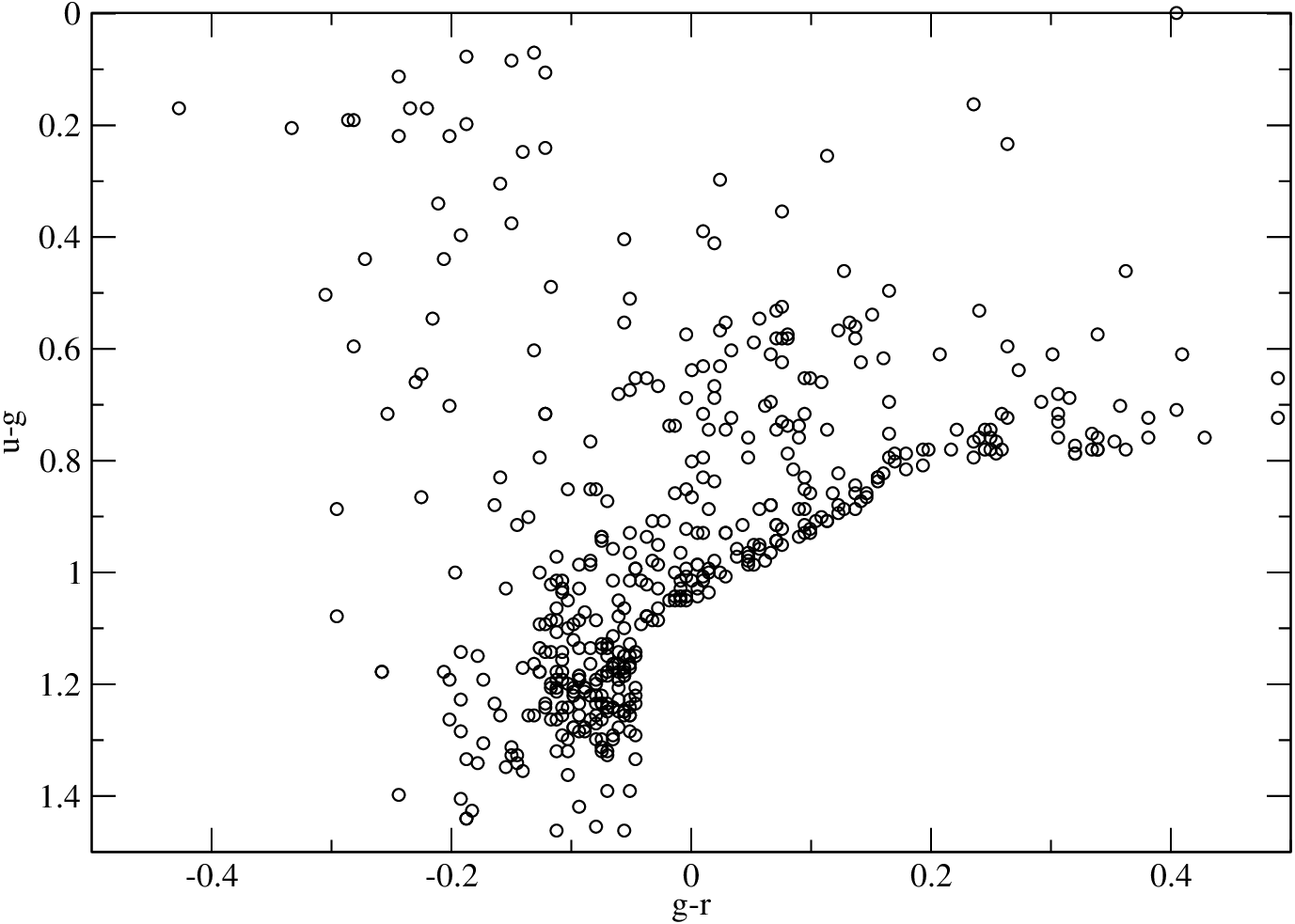}
\caption{The color distribution of BHB stars identified in the 2Qz
survey. The original Supercosmos colors have been transformed to the
SDSS using stars in common between the two surveys.}
\end{figure}

As our targets span a relatively narrow range in colors (as selected by the
2Qz survey) we assumed an absolute magnitude of $M_{b_J}=0.7 \pm 0.2$ which
is typical for BHB stars \citep{layden96}. We finally used these distances,
the known sky positions and the measured radial velocities to place all our
stars on a cylindrical coordinate system at rest with respect to the centre
of the Galaxy, assuming Solar positions as in \cite{dehnen98}. This yields a
4-dimensional (position, distance and radial velocity) map of the galactic
halo in two widely separated ($150^{\circ}$) lines of sight.

\section{The 4D Structure of the Milky Way halo}

In Figure 2 we plot the 4D map of the Galactic halo we produced, projected along
the three most relevant dimensions. It is clear from this figure that the halo
of the Milky Way extends to at least 100 kpc from the Galactic centre, and likely
well beyond (the edge of the map is set by the magnitude limit of 2Qz data),
in both directions we survey. This is considerably larger than previously
believed and comparable to the large metal-poor halo observed in the Andromeda
galaxy \citep{chapman06,kalirai06,koch08}; it would include several of the
dwarf galaxy satellites (including the Magellanic clouds) within the Galaxy's
stellar halo. As a matter of fact the Sextans dwarf is visible in Fig.~2
at $x \sim -20$ kpc and $y \sim +50$ kpc. Such large metal-poor halos may be
ubiquitous (e.g., in NGC 3379 -- \citealt{harris07}) and may be a common
byproduct of early galaxy formation.

\begin{figure*}
\includegraphics[width=6in]{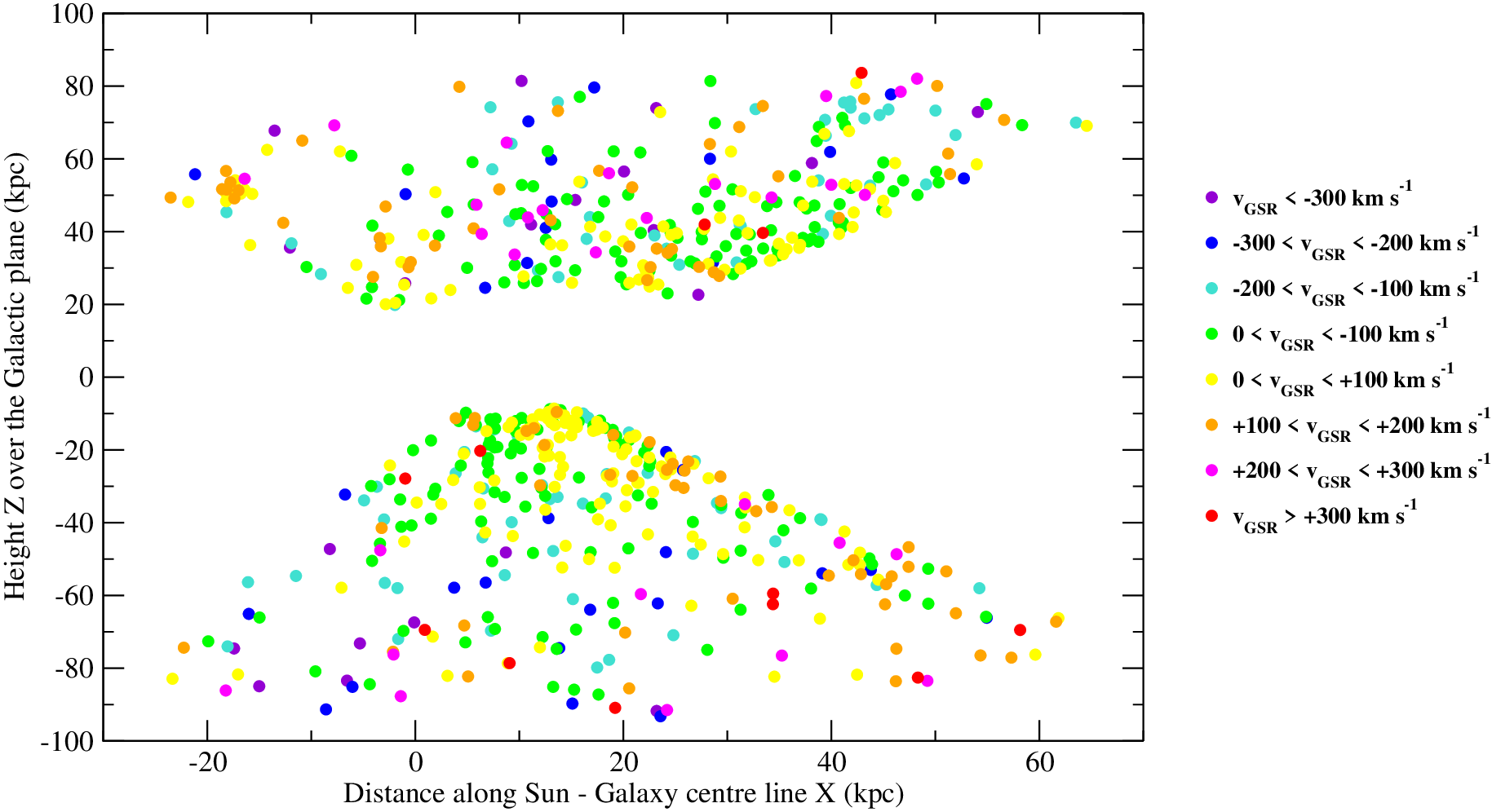}
\caption{The 4D map of the Galactic halo, with stars projected on the Galactic
plane (along a line connecting the Sun to the Galactic centre) on the x-axis and
vs. their height above the plane on the z-axis. The stars are color-coded by their
Galactocentric radial velocity, as indicated in the figure legend.}
\vspace{1cm}
\end{figure*}

In previous work, the SDSS has identified BHB stars out to 60 kpc from the
Galactic centre \citep{xue08}, while the Hypervelocity Stars Survey \citep{
brown10} found a BHB star sample to a distance of 75 kpc. The Spaghetti
survey \citep{morrison00,starkenburg09} has observed halo red giants to a
distance of 100 kpc, and star counts in the COSMOS field identify a halo
component to a distance of 80 kpc \citep{robin07}. Our study reaches
deeper than nearly all these and covers a larger field of view than
all except the SDSS (it is comparable to the Hypervelocity Stars Survey
coverage). However, we sample the BHB stars more densely as all potential 
targets have been targeted by the 2Qz survey, although of course we are
not able to classify all stars. Since we cover nearly diametrically opposite 
areas on the sky, we argue that the detection of the stellar halo in our data 
is not due to possible diffuse structures on the sky that accidentally lie in 
our line of sight (as for pencil-beam studies such as COSMOS or the Spaghetti 
survey) and the Milky Way halo truly extends to large radii.

Figure 3 shows the radial density profile of the halo along both directions we
survey. In both cases we obtain a good fit to a single power law of index $R^{-2.5
\pm 0.2}$. This is somewhat shallower than the $\sim R^{-3}$ found by \cite{morrison00}
and predicted by theory, but is in good agreement with previous measurements using
BHB stars by \cite{xue08} and \cite{brown10}. We are of course incomplete in that
we cannot detect and identify all BHB stars in 2Qz. This incompleteness is a complex
function of our ability to reliably classify stars as a function of spectroscopic
signal-to-noise. Naively, we would preferentially miss the most distant objects,
that would tend to make the radial profile steeper than it actually is, while
contamination from blue stragglers would tend to make the profile flatter. The
similarity between our profiles and those derived by (e.g.) \cite{yanny00,xue08}
and \cite{brown10} argues that our contamination fraction (from blue stragglers)
and completeness are not too different from previous samples. 

\begin{figure}
\plotone{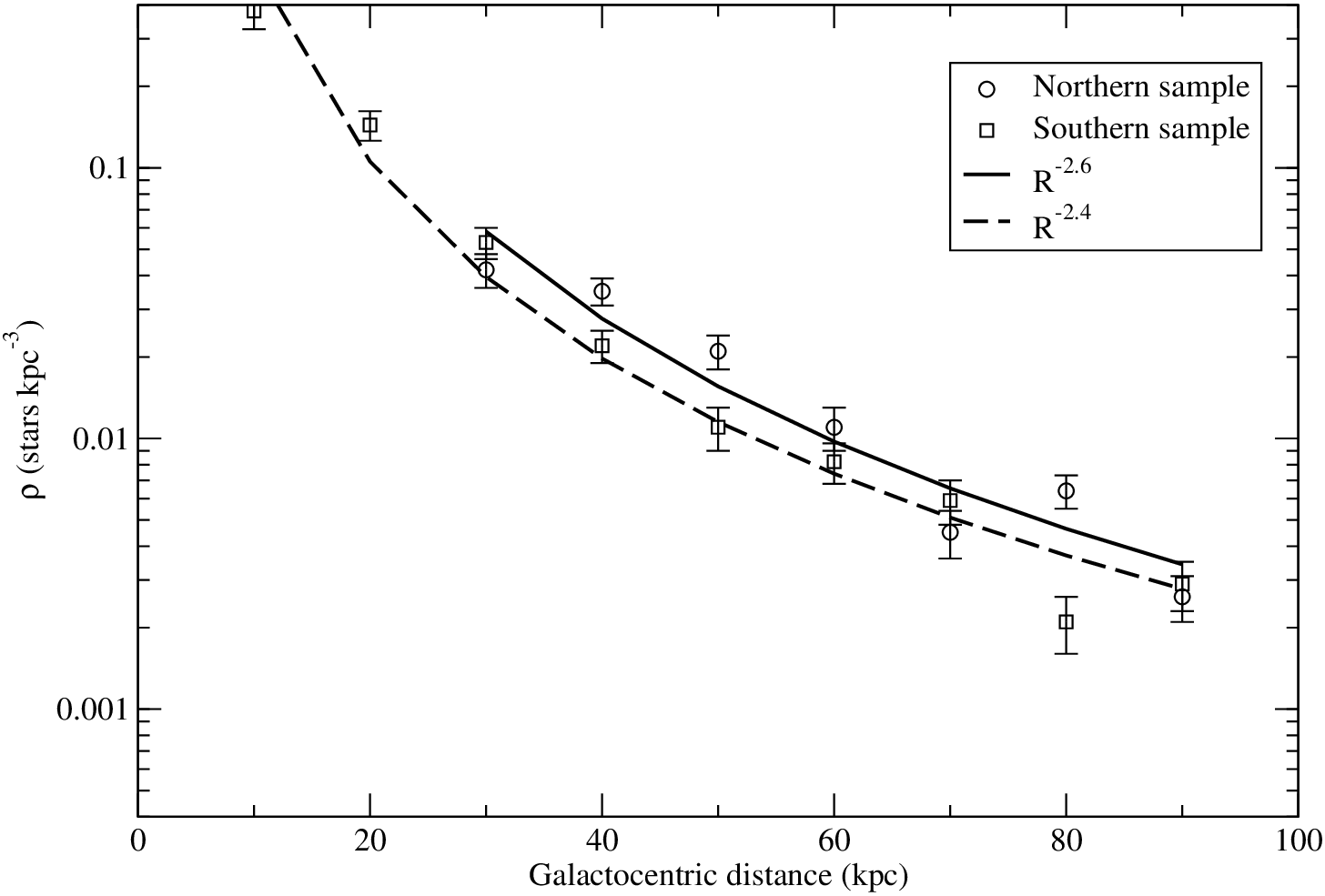}
\caption{Radial density profiles for BHB stars in the halo and best
power-law fits to the Northern and Southern samples separately. }
\end{figure}

In any case, this should not affect our differential measurement of the radial density
profile for the two individual sightlines we survey. The similarity in the profile slope
then argues that the stellar halo of the Galaxy is spherical (\citealt{majewski03}; Smith, 
Wyn Evans \& Ahn 2009), although more sightlines would be helpful to obtain a more 
precise estimate for the halo's axial ratio.

In a recent survey of RR Lyrae in SDSS stripe 82, \cite{watkins09} and \cite{sesar10}
find that the halo radial profile becomes very steep ($R^{-6}$) at galactocentric radii
larger than 40 kpc and the the outer halo seems to be dominated by a large structure
known as the Virgo-Aquila cloud, which is either an infalling dwarf galaxy or a tidal
stream. However this is not observed in our data, or those of \cite{brown10}. One
possibility is that the RR Lyrae distribution is more concentrated towards the
Galactic centre, as RR Lyrae tend to be more metal rich than BHB stars and the
halo appears to have an abundance gradient \citep{carollo07,carollo09}. This would
create an apparent break in the radial profile, reflecting the lower overall
metallicity of the sample at large distances. 

\section{Substructure in the Halo}

Inspection of Fig.~2 also shows that there are no obvious large kinematically
coherent features, such as streams, plumes or shells, in our 4D map of the
Galactic halo. We identify three possible structures: one is the Sextans
dwarf, as previously mentioned. A plume of infalling stars is present in
the top right-hand corner of Fig.~2 at $x \sim +40$ kpc and $y \sim +80$
kpc. This may be related to the Virgo-Aquila structure. Finally, a small
clump of objects is observed in the direction of Piscis Austrinus at $x \sim
+40$ kpc and $y \sim -50$ kpc and may be an undiscovered dwarf. Nevertheless,
we do not appear to observe the large streams expected in simulations \citep{
johnston08,cooper09} and the outer halo appears to be more homogeneous than
predicted.

We can quantify the presence of streams or otherwise using the Great Circle
Stream method of \cite{lyndenbell95} in Figure 4. We compute the excess radial
energy of each star compared to a smoothed Galactic potential and compare the
results to a randomized sample, where we scramble the velocities, but not the
positions, of stars, 1000 times. Streams would show up in this figure as long 
lines of stars. It is easy to see that our sample contains no significant stellar 
streams. The large metal poor outer halo appears to represent an extension of the 
smooth metal poor inner halo structure \citep{carollo07,carollo09,bell08}.

\begin{figure}
\plotone{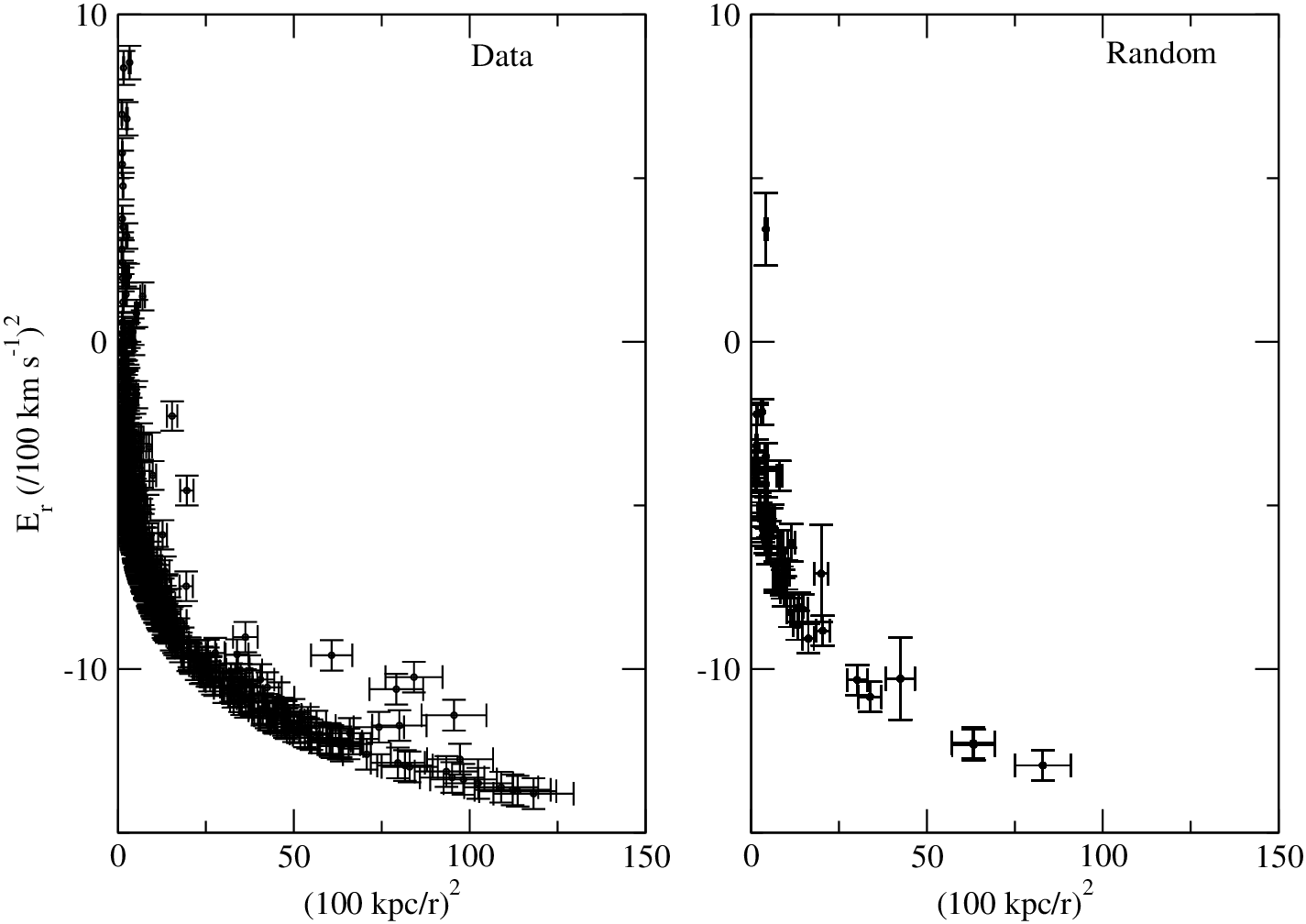}
\caption{The Great Circle Stream method applied to stars in our sample.
We plot the excess radial energy of each stars with respect to a smoothed
Galactic potential vs. its radial distance from the centre of the Galaxy.
The left hand panel shows the actual data, where the right hand panel
is a randomization. No streams are observed in these figures as long
lines of stars having the same excess radial energy.}
\vspace{1cm}
\end{figure}

However, lack of large streams does not mean lack of substructure. \cite{bell08}
find that about 40\% of stars in the inner halo are substructured, even if the
only prominent stream in their data is the well known Sagittarius stream, while
\cite{starkenburg09} find that around 25\% of the stars in the Spaghetti survey
are more clustered than expected from a random distribution. We apply the 
4-distance method used by \cite{starkenburg09} to our data in Figure 5. This
is essentially a version of the correlation function for 4-dimensional data,
computing the excess number of stellar pairs in position and velocity compared
to a random sample (realized by scrambling the velocities but not the positions
of the stars in the data) as a function of distance in 4-dimensional space.
The angular, spatial and velocity separations are scaled by the range in these
quantities covered by the data.

\begin{figure}
\plotone{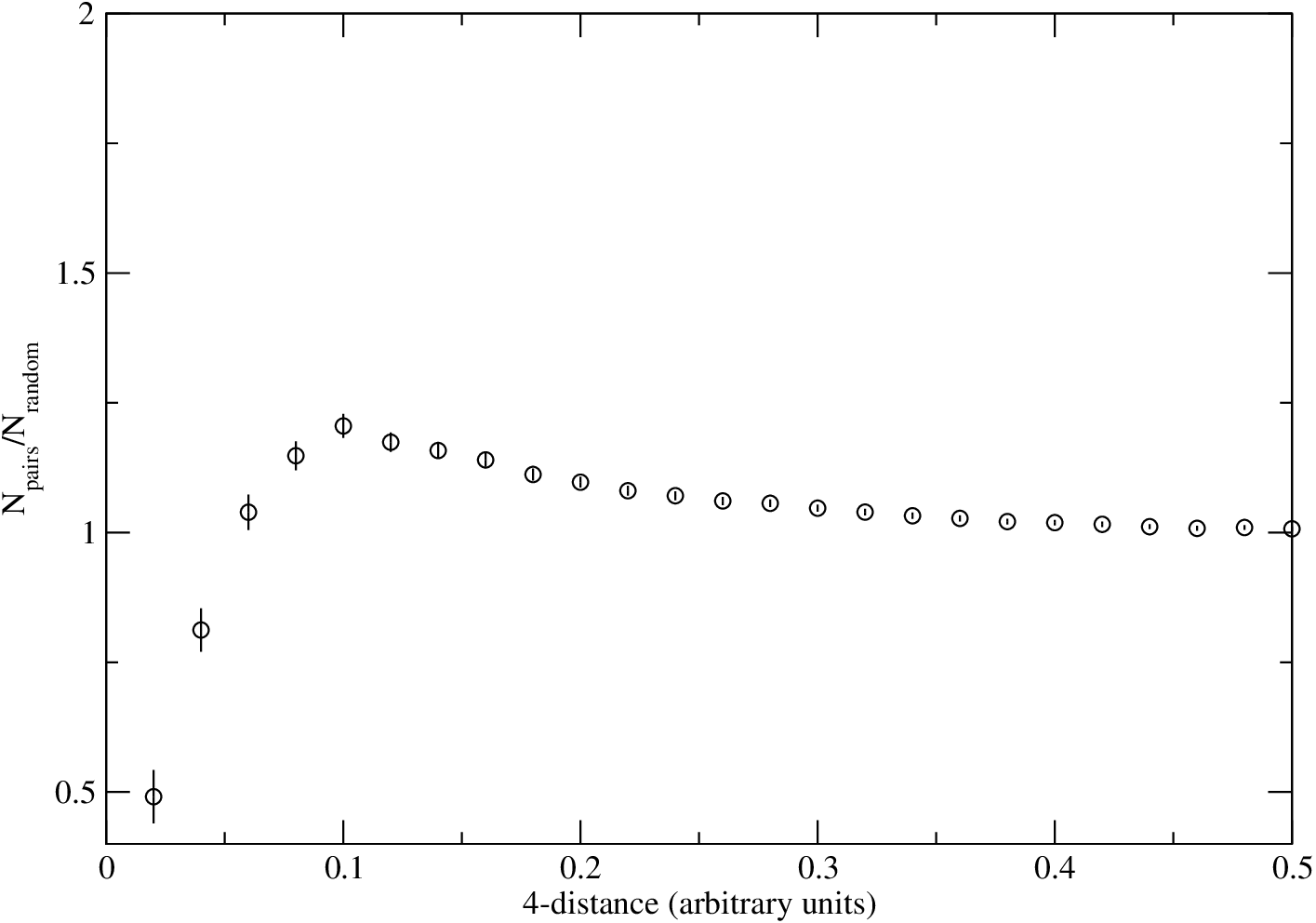}
\caption{The excess fraction of stellar pairs over random (see text)
in 4-dimensional distance for our sample of BHB stars.}
\end{figure}

At a 5$\sigma$ level we find that at least 20\% of our stars are more paired 
than random. This agrees, broadly, with the estimates by \cite{bell08} for the
inner halo and \cite{starkenburg09} for the outer halo. In addition we detect
a decrease in the correlation strength at small 4-distance. This is expected
if the outer halo is dynamically young and suggests the presence of numerous
streamlets and a complex structure, too weak to be resolved by our data. We
present further evidence to this effect from an analysis of halo kinematics.

\section{Kinematics of the Halo}

The kinematics of metal-poor stars in the halo yield information on the earliest
phases of galaxy formation \citep{eggen62} and the shape of the Galactic potential.
Until recently, known samples of halo stars were small, especially at large 
galactocentric distances. \cite{battaglia05} used a heterogeneous sample of
red giants, BHB stars, globular clusters and dwarf galaxies to trace the velocity
dispersion profile of the Galaxy out to $\sim 100$ kpc, finding a mildly declining
profile. Using BHB stars in SDSS \cite{xue08} found a flat or mildly declining
profile out to 60 kpc and therefore inferred a large mass of the Milky Way. 
\cite{brown10} also derive a mildly declining profile out to 75 kpc using BHB
stars in the Hypervelocity Star Survey. In all these cases, the fits depends
strongly on the accuracy of the last (most distant) data points, which generally
contain fewer objects.

We use our data to derive the velocity dispersion profile in both Northern
and Southern samples separately. We use bins containing equal numbers of
stars and calculate the velocity dispersion and its error following a 
maximum likelihood approach \citep{walker06}. Figure 6(a,b) plots our results.
While we are in reasonable agreement with previous work over the range
where we overlap, we find evidence of a rising velocity dispersion at
large radii, in both fields we study. This is unlike the results of
\cite{battaglia05} and \cite{brown10}, although we reach farther into
the distant halo than they do. In the former case, the heterogeneous
sample and its relatively small size may cause part of the difference,
as it is known that different tracers have different kinematics (e.g.,
\citealt{kinman07}). In the latter case, we and \cite{brown10} use
the same tracers, but while our stars are concentrated in two narrow
strips, \cite{brown10} uses a wide region selected from the SDSS.
We have more stars (by a factor of about 3) than they do at large 
($R > 50$ kpc) distances. One possibility is that we sample this
regime more densely, especially if the halo is as inhomogeneous as
theory suggests and as the evidence from Fig.~5 also argues.

\begin{figure}
\plotone{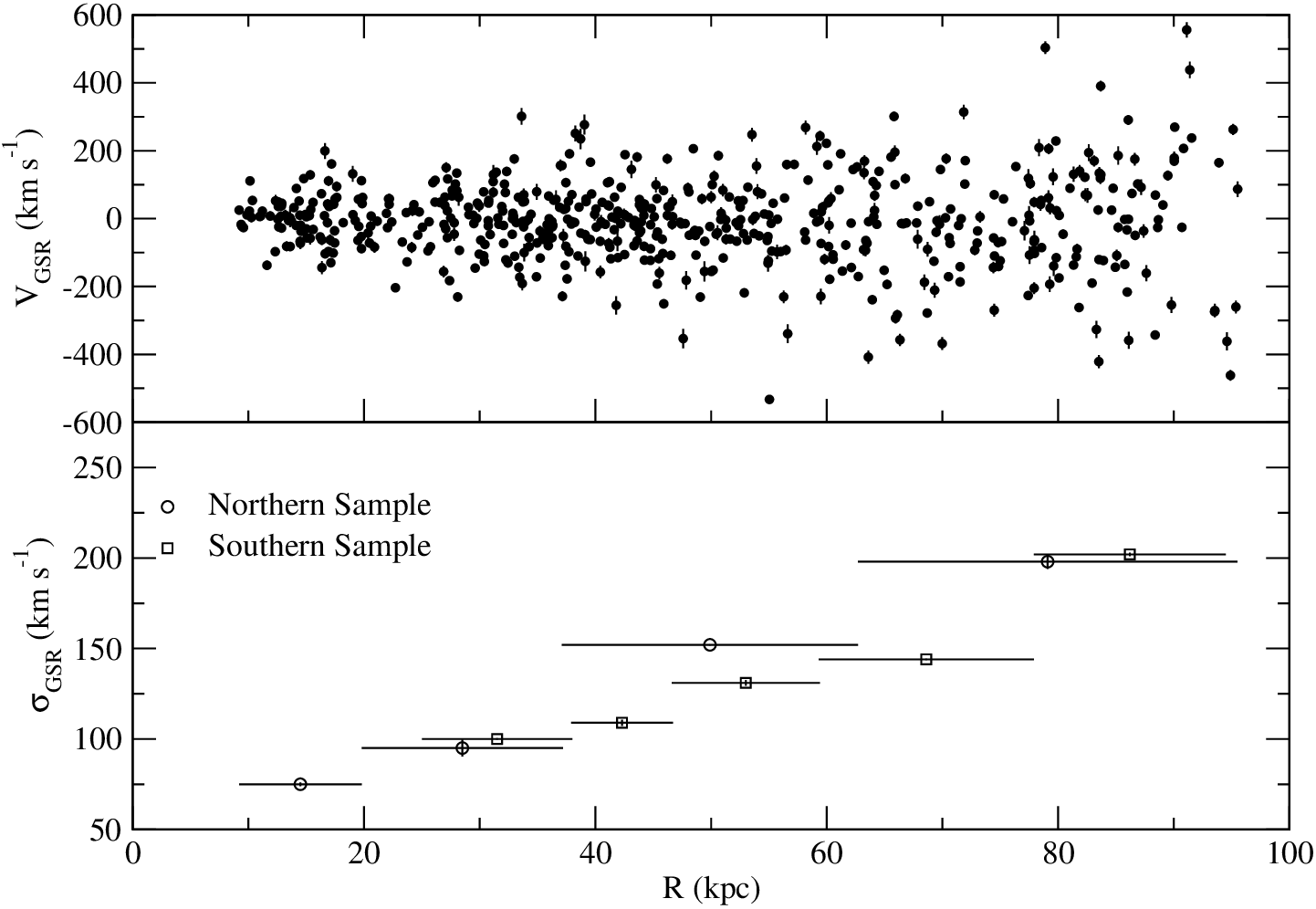}
\caption{{\it Top panel:} Run of radial velocities of BHB stars as a function
of Galactocentric distance. {\it Bottom panel:} The radial velocity dispersion 
profile of the Galactic halo in both Northern and Southern samples. Each bin
contains equal numbers of stars and the velocity dispersion and its error are
calculated using a maximum likelihood method. The error bars on points for the 
x-axis represent the size of the bin used.}
\end{figure}

Is it possible that our result may be affected by contamination from
blue straggler stars. Since these objects are 2 to 3 magnitudes fainter
than BHB stars, blue stragglers will contaminate the bins at $R_{GC} 
> 40$ kpc with objects that truly lie at distances of 10 to 30 kpc.
The selection procedures we describe above should exclude most contaminants
but even if we are as effective as previous studies, $\sim 20\%$ of our
sample may consist of blue stragglers. In order to assess the effects
of contamination, we have used a model distribution function for the 
velocity dispersion profile of our Galaxy from van Hese, Baes \& Dejonghe
(2009), with Galactic mass of 1.1 $\times 10^{13}$ M$_{\odot}$ and halo
scale radius of 40.5 kpc from \cite{dehnen98}. We sampled 100 stars at
distances of 10 to 100 kpc with a Gaussian random deviate having $\sigma (r)$
from \cite{vanhese09} and zero mean velocity. We adopted two cases: one 
with no velocity anisotropy $\beta_0=\beta_{\inf}=0$ and one with fully 
tangential radial velocities at infinity ($\beta_0=0$; $\beta_{\inf}=1$). 
For each of these we adopted a variable contamination, between 10\% and 40\%, 
from blue straggler stars, which we assumed to be 2.5 mag. fainter than BHB 
stars. We replaced this fraction of stars in each bin with $R_{GC} > 40$ kpc, 
with blue stragglers at the appropriate distance, with velocities sampled
from a Gaussian random deviate with $\sigma (r)$ from \cite{vanhese09},
at the appropriate $r$ for a contaminating blue straggler. The results of this 
exercise are shown in Figure 7. Although the blue stragglers increase the
measured velocity dispersion at large radii (compared with the theoretical
profile), they do not produce a flat or rising velocity dispersion profile (as in
the bottom panel of Fig.~6) even at 40\% contamination, and even for the 
relatively flat $\sigma(r)$ profile produced by the  fully isotropic model.

\begin{figure}
\plotone{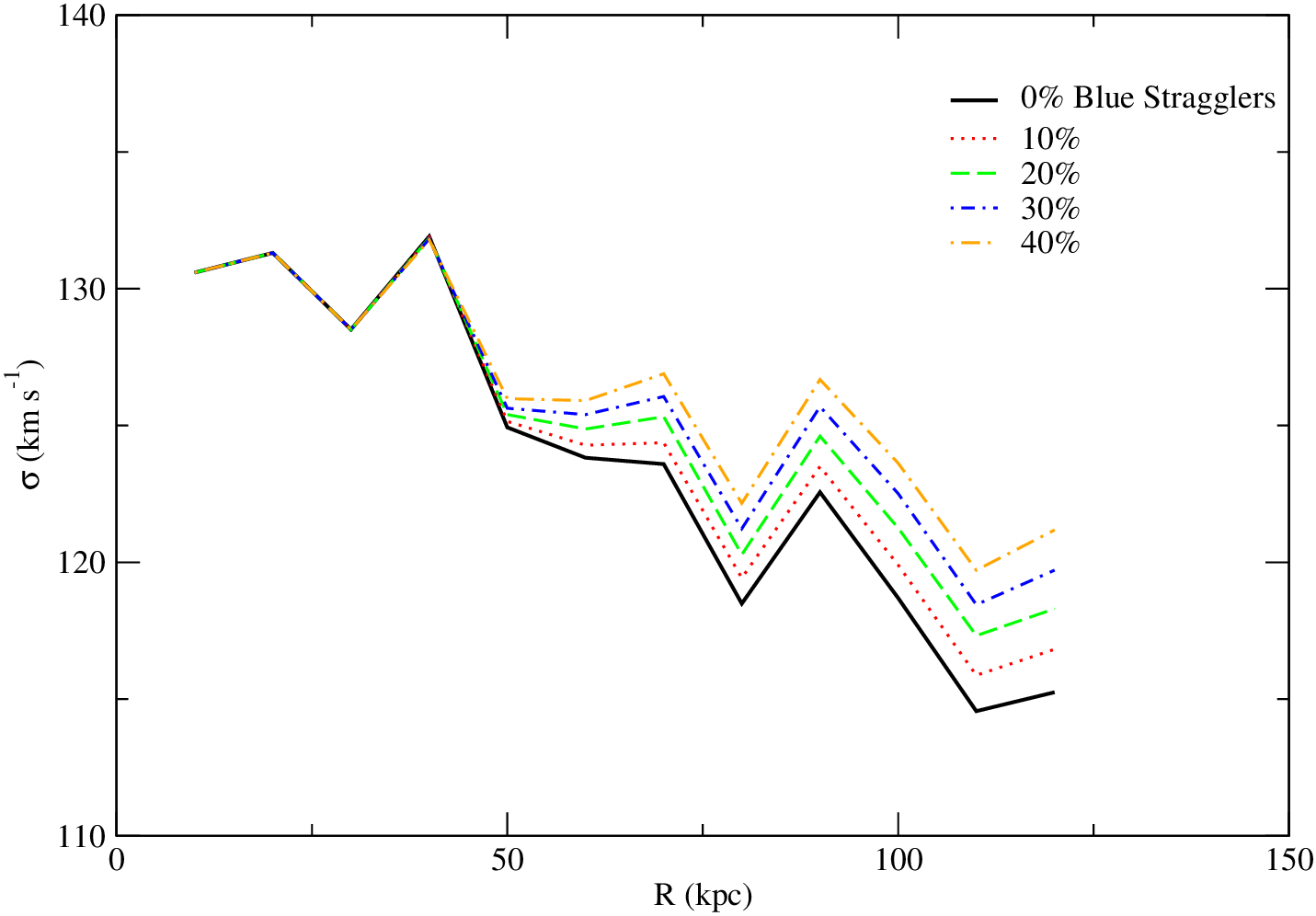}\\
\vspace{1cm}
\plotone{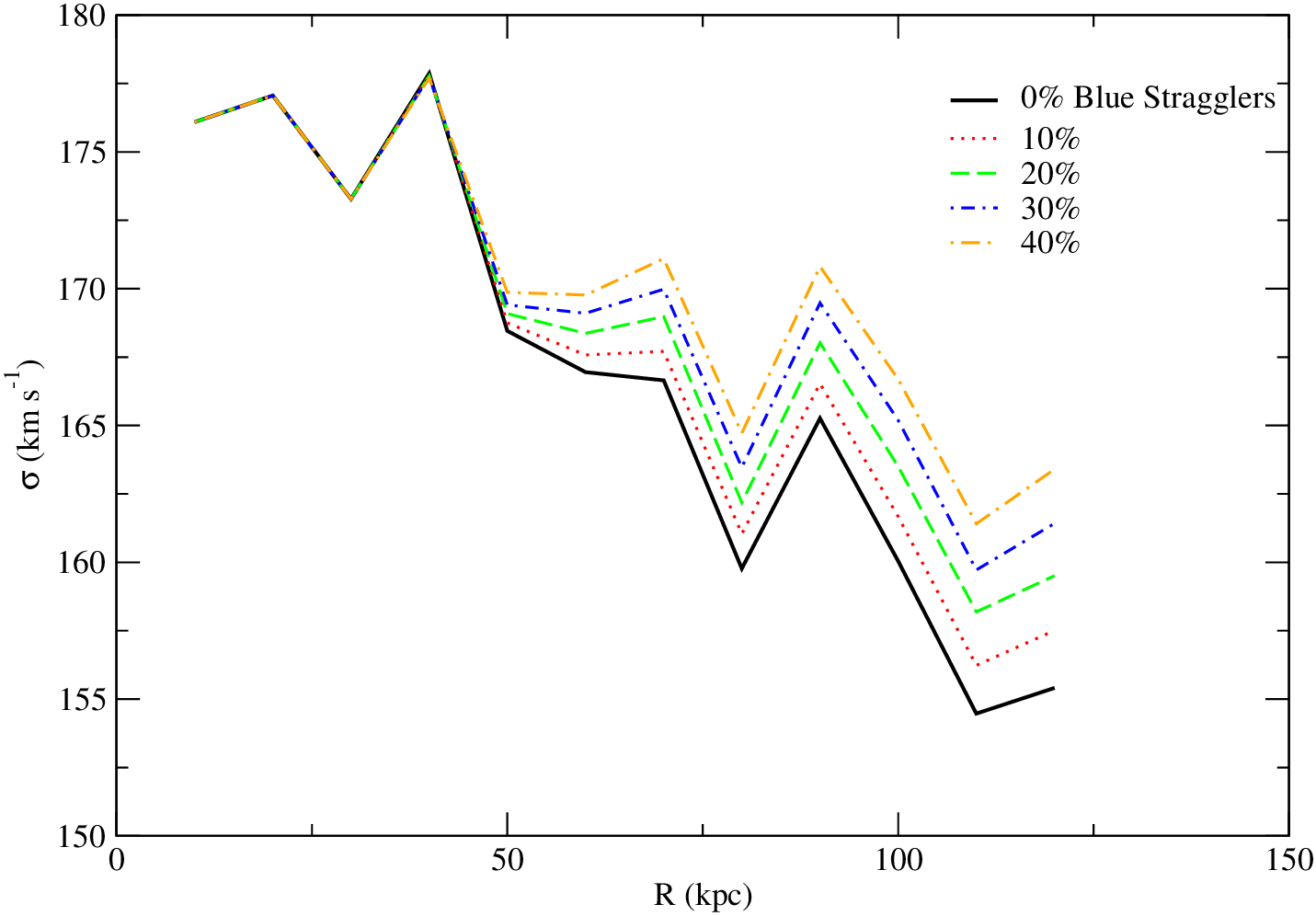}
\caption{The effect of BHB star contamination on the velocity dispersion
profile of the Galaxy. The top panel is for a distribution with no
velocity anisotropy while the bottom panel assumes that all velocities
are tangential at large radii. See text for details. The degree of 
contamination assumed is indicated in the legend.}
\end{figure}

In Figure 8 we plot the histograms of radial velocities for all stars
in 6 bins, containing equal numbers of objects and covering a range of
distances. We see that while the distributions are acceptably Gaussian
within the inner 60 kpc, we observe both an increase in dispersion and
a loss of Gaussianity in the two outer bins. This is consistent with
what we observe in Fig.~5 and 6, where we find indications of an increasing
velocity dispersion and a dynamically young and substructured halo. If
the outer halo consists of a myriad of streamlets, accreted more or less
recently from the disruption of satellites or low luminosity galaxies,
we would expect evidence of dynamical youth and tidal heating, as we
observe in Fig.~4 and 5.

\begin{figure}
\vspace{1cm}
\plotone{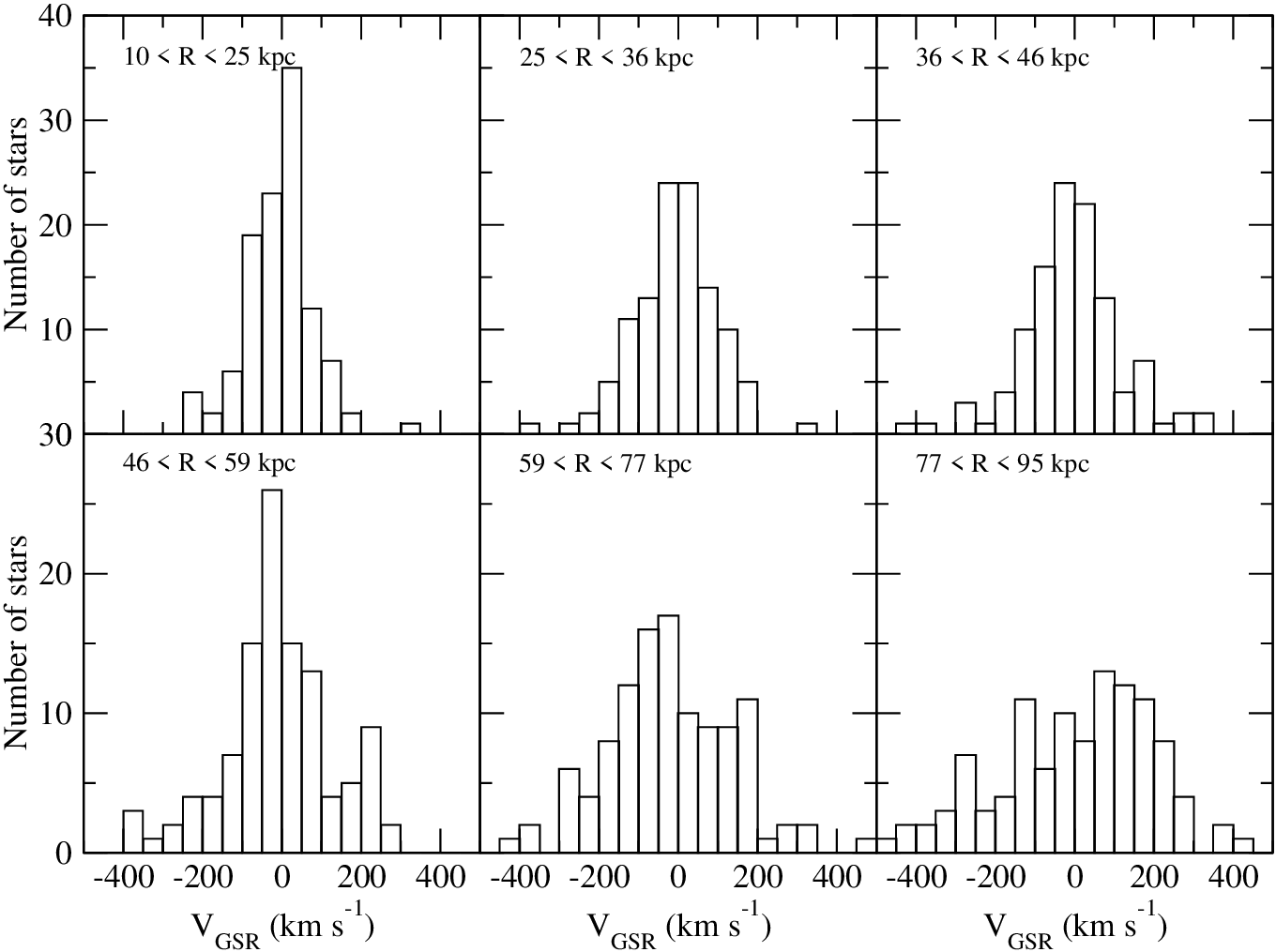}
\caption{Histograms of the velocity distribution for 6 bins containing
equal numbers of stars (the distance ranges sampled are indicated in
figure legends). The two more distant bins appear to have a much less
Gaussian distribution than the four bins at $R < 60$ kpc.}
\end{figure}

A possible alternative explanation is that the dark halo is very
massive. \cite{gnedin10} has recently argued for a very massive 
halo based on the flatness of the velocity dispersion profile. 
However, in this case the mass to light ratio needed would 
exceed several hundred. Tidal heating may also produce an increasing
velocity dispersion, as is observed in some dwarf galaxies (Munoz,
Majewski \& Johnston 2006). These explanations, while possible,
would not probably produce the non Gaussian velocity dispersion
observed at large radii.

The combined evidence from the space distribution and kinematics of
BHB stars points to a very large and metal poor halo, whose outer
regions appear to have been accreted relatively recently from low
luminosity satellites, analogous to the copious tidal debris 
observed in M31. Although large streams are not observed the
data appear to be in comparatively good agreement with theoretical
predictions, although it is possible that minor mergers are more
important than expected. More observations as well as models to
truly `observe' the simulated halos in the same fashion as the
data and directly compare theory and reality will be needed to
refine our understanding of the formation of the Milky Way.

\acknowledgments

We would like to thank the anonymous referee for a very helpful report which
clarified many points in this paper. We thank the 2dF Quasar Survey for making
their data available to the community and in particular Scott Croom for answering
many questions on the data. We warmly thank all the present and former staff of 
the Anglo-Australian Observatory for their work in building and operating the 2dF 
and 6dF facilities. The 2QZ and 6QZ are based on observations made with the AAT 
and the UKST. 



{\it Facilities:} \facility{AAT (2dF)}.

\end{document}